\documentclass{article}
\usepackage{spconf,amsmath,graphicx,xcolor}
\usepackage{subcaption}
\usepackage{multirow}
\usepackage{tabularx}
\usepackage[colorlinks = true,linkcolor = blue,urlcolor = blue]{hyperref}
\newcommand{\C}{\boldsymbol{C}}
\newcommand{\W}{\boldsymbol{W}}
\newcommand{\B}{\boldsymbol{B}}
\newcommand{\Sb}{\boldsymbol{S}}

\newcommand{\x}{\boldsymbol{x}}
\newcommand{\y}{\boldsymbol{y}}

\def\x{{\mathbf x}}

\title{Audio Retrieval for Multimodal Design Documents: A New Dataset and Algorithms}
\name{Prachi Singh$^{1,2}$\thanks{This work was done as part of internship in Adobe Research.}
, Srikrishna Karanam$^1$, Sumit Shekhar$^1$} 
\address{
$^1$Adobe Research, Bangalore, India \\
$^2$Indian Institute of Science, Bangalore, India \\
\texttt{\small{prachisingh@iisc.ac.in}}}
\begin{document}
\ninept
\maketitle
\begin{abstract}
We consider and propose a new problem of retrieving audio files relevant to multimodal design document inputs comprising both textual elements and visual imagery, e.g., birthday/greeting cards. In addition to enhancing user experience, integrating audio that matches the theme/style of these inputs also helps improve the accessibility of these documents (e.g., visually impaired people can listen to the audio instead).

While recent work in audio retrieval exists, these methods and datasets are targeted explicitly towards natural images. However, our problem considers multimodal design documents (created by users using creative software) substantially different from a naturally clicked photograph. To this end, our first contribution is collecting and curating a new large-scale dataset called Melodic-Design (or MELON), comprising design documents representing various styles, themes, templates, illustrations, etc., paired with music audio. Given our paired image-text-audio dataset, our next contribution is a novel multimodal cross-attention audio retrieval (MMCAR) algorithm  that enables training neural networks to learn a common shared feature space across image, text, and audio dimensions. We use these learned features to demonstrate that our method outperforms existing state-of-the-art methods and produce a new reference benchmark for the research community on our new dataset.

%Multimodal design document consists  of images, illustrations, style elements along with text which helps in effective communication of the content e.g. birthday card. Integrating audio in the form of music in the design documents will further enhance user experience. This will also help in increasing the accessibility to a wider audience.  Since very less work is done in this area, there was no open source  dataset available to perform this task. 
%Even though task is important, very less work is done in this area. 
%Therefore, in this paper we propose our own dataset for multimodal music retrieval called as "Melodic-Design" dataset which consists of images containing different styles, templates , illustrations and corresponding caption mapped to a set of music audios based on some common moods/themes. Based on this dataset, we perform music retrieval task given an image-text pair. For benchmarking, existing models which perform song retrieval based on human centric images are used. Additionally, we propose two branch music retrieval model which computes the similarity between image-caption and audio to retrieve k best music suggestions.
%Our proposed work is an attempt to open the doors of opportunities to explore and provide solutions in this direction. This paper discusses  performs music retrieval from an . Existing    
\end{abstract}
\begin{keywords}
Music Retrieval, Multimodal processing, cross attention. 
\end{keywords}
\section{Introduction}
\label{sec:intro}

\begin{figure}[htb]
\centering
\includegraphics[width=0.4\textwidth,trim={5.5cm 4.9cm 8.3cm 4.3cm},clip]{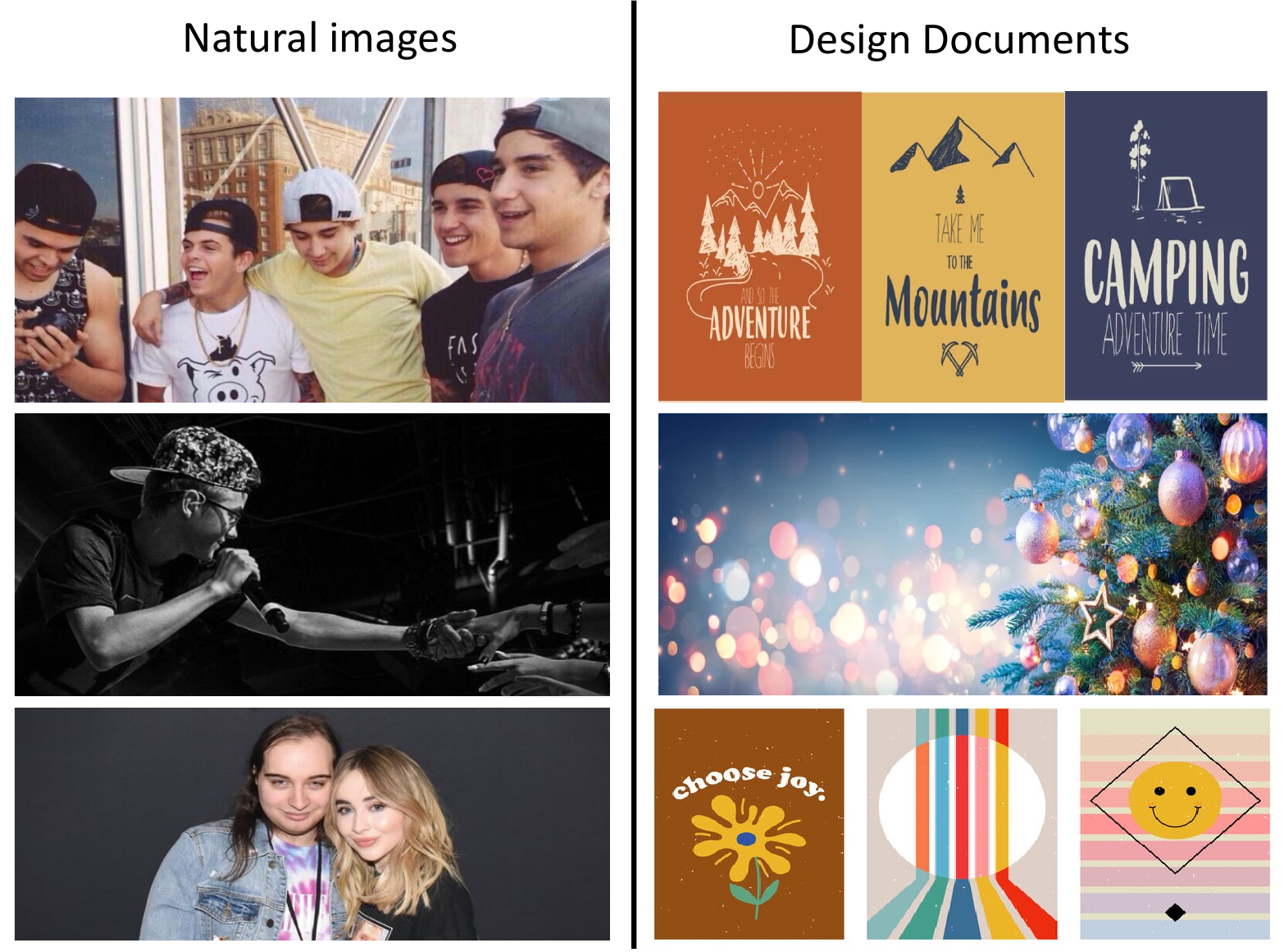}
\caption{Natural images (left) vs. design documents (right).}
\label{fig:imagedesign}
\end{figure}

With increasing proliferation of on-demand web/mobile-based graphics design softwares\footnote{https://www.adobe.com/express}$^{,}$ \footnote{https://www.canva.com}$^{,}$ \footnote{https://www.sketch.com}, 
designing creative documents has become very easy for any occasion, e.g., greeting cards, event invitations/flyers, social media infographics etc. In most cases, such design documents tend to be multimodal, i.e., they comprise some visual imagery aspects and some textual elements (see Fig~\ref{fig:imagedesign}(right)). For such documents, adding an additional modality in the form of relevant audio/music files will not only enhance the consumption experience of users but also improve document accessibility for visually impaired users. To this end, our first contribution is the consideration and proposal of a new problem involving the retrieval of relevant audio files given a multimodal design document. While much work in the past \cite{liang2018jtav,chen2020vggsound} has focused on audio retrieval for natural images, there has not been any work in the context of the kind of design documents referred to above, and this paper takes a step towards bridging this gap in the literature (see Fig~\ref{fig:imagedesign}). 

%As noted above, since much recent work \cite{liang2018jtav,chen2020vggsound,wu2022wav2clip} focuses on audio retrieval given natural images, the datasets they use also contain only natural images and are not suitable for our proposed problem. 
As the problem is unexplored, the existing datasets \cite{li2017image2song,chen2020vggsound}  for audio retrieval contains only natural images and are not suitable for our proposed problem.
To this end, our second contribution is the collection and curation of a new paired design-audio dataset that comprises multimodal design documents, scraped from publicly available data from Adobe Stock \footnote{https://stock.adobe.com/}, paired with relevant audio files collected from the MTG-Jamendo \cite{bogdanov2019mtg} repository. With $\approx 500k$ design documents paired with over $\approx 7.5k$ audio files, this is a first-of-its-kind dataset that we believe will help advance research in multimodal design understanding. 

Finally, our third contribution is a novel multimodal cross-attention algorithm that enables training neural networks to learn a shared representation among the image, text, and audio modalities present in our problem setting. In particular, given paired design-audio samples, we extract individual modality features and learn per-pair as well as overall weights to learn a unified design-audio embedding. With extensive experiments on our proposed new dataset, we demonstrate our algorithm substantially outperforms the existing state-of-the-art audio retrieval methods.

%Audio Retrieval is the task of searching for audio files similar to the input query. This is done for different purposes like automatic text to speech conversion, audio segmentation, environmental sound retrieval, music information retrieval.  Nowadays, documents are a very important means of communication. Therefore, it not only contains text but other various components like designs, infographics, animations etc. for easy and effective access to information which can be referred as multimodal design document. Document consumption can further be made more exciting by adding audio modality to it. 
%In our work, we focus on the task of retrieving music audio for design documents containing images with different style elements e.g templates, illustrations, artwork etc.  and text to make them more interactive and engaging.

%The proposed work aims to integrate music in the document to further enhance user experience increase accessibility for all. 

\section{Related Work}
\label{sec:rel_work}
%As noted in Section~\ref{sec:intro}, much work exists in multimodal audio retrieval, but these methods focus on natural images. 
As noted in Section~\ref{sec:intro}, works in multimodal audio retrieval are mainly focussed on natural images. 
In particular, in Image2Song \cite{li2017image2song}, the shuttersong dataset was used to map images and song lyrics to the same feature space, which was then used for downstream tasks like retrieval. Using the same dataset, Liang et al. \cite{liang2018jtav} proposed a method to jointly learn a feature space by fusing visual and acoustic features. Similarly, even datasets like VGGSound \cite{chen2020vggsound}, and MUGEN \cite{hayes2022mugen}, while having a video modality, also focus on either naturally occurring human actions or videos generated by game engines. In contrast, our contribution is unique by proposing a new dataset solely focused on multimodal creative design documents like greeting cards, infographics etc., that are commonly created using creative design software. Our large-scale dataset comprising hundreds of thousands of design documents paired with audio files provides a challenging testbed for advancing retrieval research.

\section{Melodic Design (MELON) - A new dataset}
\label{sec:dataset}

%However, the existing datasets are mostly focused on human-centric images e.g photographs of people and events. Therefore, we created our own dataset called as Melodic-Design dataset.  The details of dataset creation and statistics are discussed in the following sections.

As discussed in Section~\ref{sec:intro}, given a multimodal design document like the ones shown in Fig~\ref{fig:imagedesign}, our problem is one of retrieving a short list of audio files that \textit{go well} with the various elements of the input. For example, the first row/second column in Fig~\ref{fig:imagedesign} shows an adventure-themed design document containing both images and text. Different elements include the background image/color, text fields (e.g., ``Mountains"), and decorative elements and shapes. Note that each of these elements forms a \textit{layer} in the design document (e.g., background image is the background layer, and the textual greeting is the foreground layer), giving a multi-layered multimodal design document. As noted in Section~\ref{sec:intro}, and also from Figure~\ref{fig:imagedesign}, existing datasets focus solely on natural images, whereas our problem entails design documents. To bridge this clear gap in the literature, we collect and curate a new dataset comprising pairs of multimodal design documents and corresponding audio files, and we call our dataset ``MELOdic desigN" (MELON). %We next discuss our data collection procedure and the various salient features of the dataset. 

\begin{figure}
\centering
\includegraphics[width=0.4\textwidth,height=5cm]{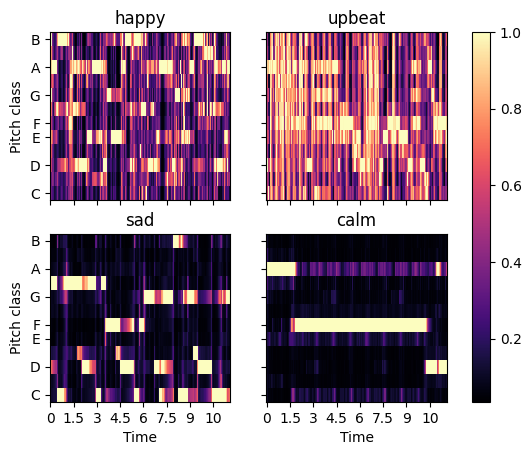}
\caption{Comparison of time-frequency chroma features of audio of happy, sad, upbeat and calm moods. Colour (dark to light) represent the energy content (low to high) in each time-frequency bin. }
\label{fig:chromaImg}
\end{figure}

\begin{table}[t]
\scalebox{0.8}{
\begin{tabular}{l|llllll}
\hline
Dataset                                                                  & \begin{tabular}[c]{@{}l@{}}Visual \\ Content\end{tabular} & Modalities  & \#Images & \#Audio &  Tasks \\ \hline
VGGSound \cite{chen2020vggsound}                                   & Action vid.                                            & I+A                                                                  & 199k     & 199k    & C \\ 
Audio set \cite{gemmeke2017audio}                                     & Human vid.                                          & I+ A                                                                 & 2.1m         &2.1m         & R + G       \\
MUGEN \cite{hayes2022mugen}                                        & Game vid.                                           & I+A+T                                                             & 233K     &233K         & R+G        \\
Shuttersong \cite{li2017image2song}                                & Images                                              & I+A+T                                                           & 17k      & 17k     & MR + C  \\
IMEMNet \cite{zhao2020emotion}                                   & Images                                              & I+A                                                                  & 25k      &1.8k         & MR   \\ \hline
\begin{tabular}[c]{@{}l@{}}\textbf{MELON} \\ {[}Proposed{]}\end{tabular} & Various                                                 & I+A+T                                                            & 488k     &7.7k         & MR + C \\ \hline
\end{tabular}
}
\caption{Melodic-Design vs. other datasets. I, A, and T corresponds to image, audio, and text respectively.  ``Various" for Melodic-Design covers illustrations, vectors, template, and background designs. C, R, G, MR represents classification, retrieval, generation, and music retrieval respectively.}
\label{tab:datasets}
\end{table}

%\begin{figure}[htb]
%\includegraphics[width=0.5\textwidth]{figures/plot_sub.png%} 
%\caption{Histogram plot of top 20 moods based on number of %audio samples present in the mood.}
%\label{fig:moodaudio}
%\end{figure}

%\begin{figure}[htb]
%\includegraphics[width=0.5\textwidth,trim={1.5cm 3cm 3.5cm %3cm},clip]{figures/mood_wordcloud_v2.pdf}
%\caption{The figure shows word cloud for 9 different mood %categories in the proposed dataset. Each subplot shows %words present in the mood extracted from the captions with %varying size and colour. The size is representative of the %frequency of the word in the captions corresponding to %that mood.}
%\label{fig:wordcloud}
%\end{figure}

%\begin{figure}
%  \begin{subfigure}{0.25\textwidth}
%    \includegraphics[width=\linewidth]{figures/happy_sad_chroma.png}
%    \caption{Moods: happy, sad} \label{fig:1a}
%  \end{subfigure}%
%  \hspace*{\fill}   % maximize separation between the subfigures
%  \begin{subfigure}{0.25\textwidth}
%    \includegraphics[width=\linewidth]{figures/upbeat_calm_chroma.png}
%    \caption{Moods: upbeat, calm} \label{fig:1b}
%  \end{subfigure}%
%  \caption{Comparison of time-frequency chroma features of audio of happy, sad , upbeat and calm moods. }
%  \end{figure}

\subsection{Collecting Raw Dataset Samples}
We use the publicly available MTG-Jamendo \cite{bogdanov2019mtg} database that encompasses a variety of mood/theme categories, instruments, and genres as our source of audio files. We use various time-frequency features like intensity, timbre, pitch, tempo, and rhythm \cite{bhat2014efficient} to identify the mood of audio. For example, the pitch varies from very high to very low as we move from the ``happy" to ``sad" mood. Similarly, the intensity and tempo of mood ``upbeat" is very high, whereas that of ``calm" is very low (see Fig.~\ref{fig:chromaImg}). For mapping/associating audio to design documents below, we use music files corresponding to 50 mood categories in MTG-Jamendo. Since MTG-Jamendo also has audio files labelled with multiple mood categories, we only retain those data samples labelled with only one mood for simplicity.

We use publicly available data from Adobe Stock as our source for collecting multimodal design documents comprising image and text content. To scrape images, we built a software utility that can query Stock with any mood category along with data types as part of the input. For instance, one such query would involve fetching illustrations, vectors, templates, and background images for the \texttt{adventure} mood. %This helps increase the variability of the dataset and not limit to only one type of documents (e.g., background images). 
By restricting the search-page limit to 10, we obtain about $10,000$ images across all the above document types for every mood category. 
%Note that we can also obtain the caption for the image as part of this scraping process, resulting in 10k image-text pairs for each mood category. 
Note that Adobe Stock also provides image metadata which contains manually generated captions describing the image elements in detail. % e.g. caption for Figure \ref{fig:designdoc} is \textit{``Christmas background with branches and holiday balls."} 
%\snote{add one line how these captions were generated in Stock, e.g., was it manually captioned?}  

%The keywords in the captions are used to describe the variability of the image dataset in the statistics section (Section \ref{sec:datastat}). 

\begin{figure*}
\centering
\includegraphics[width=12cm, trim={0cm 2cm 0cm 1.5cm},clip]{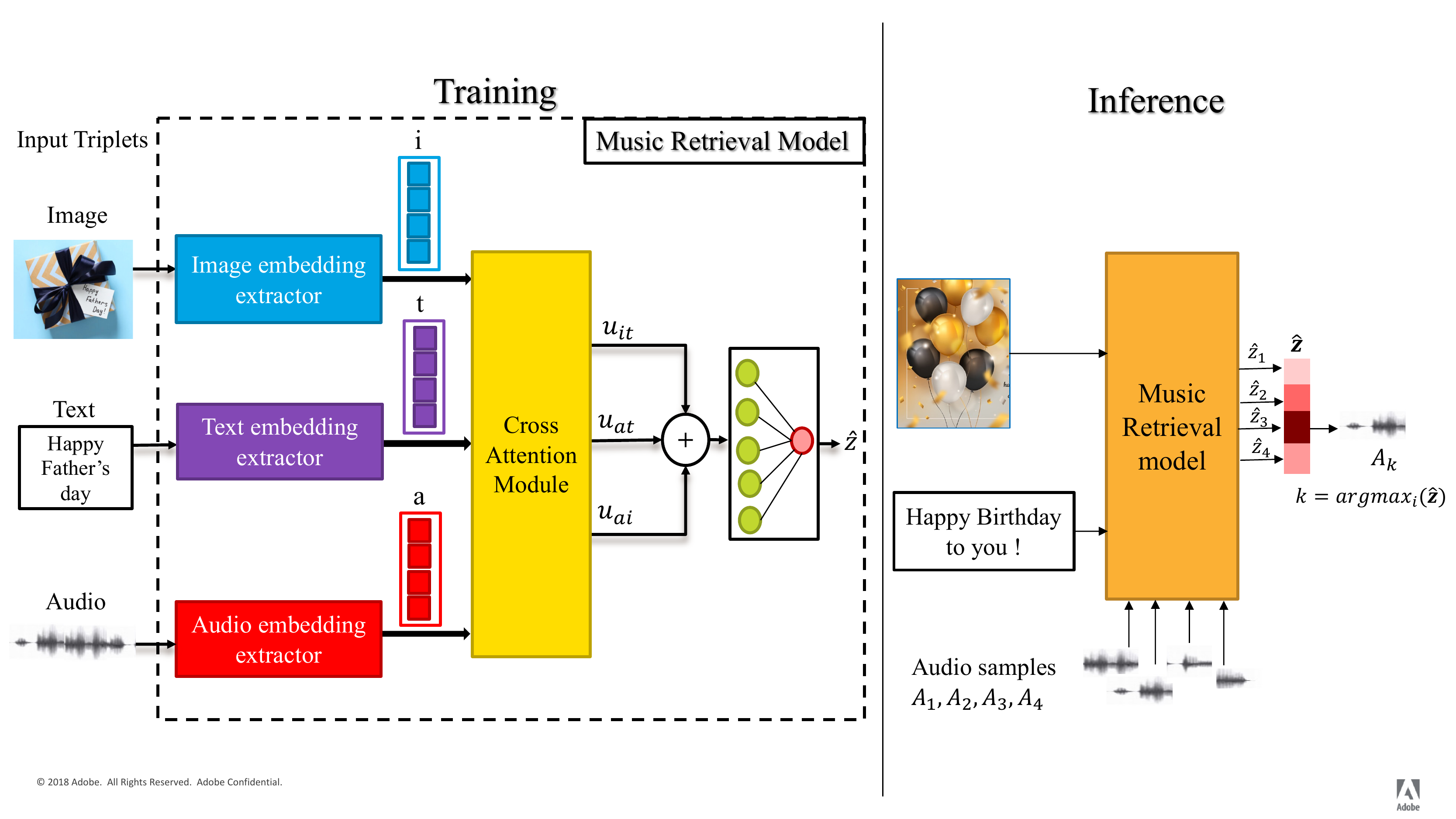} 
\caption{Block diagram of the proposed music retrieval model using image and text pair.}
\label{fig:blockdiagram}
\end{figure*}

\subsection{Establishing Correspondence \& Dataset Statistics}
\label{sec:pairformation}
In our dataset, images and text are already paired since each downloaded image comes with a ground-truth caption. We use the common mood categories across the MTG music dataset and our proposed design document dataset to form image-caption-audio pairs. Specifically, given a mood category, we first extract the CLIP \cite{radford2021clip} features for an image-text sample. For each audio file tagged with the same mood, we extract Wav2CLIP \cite{wu2022wav2clip} embeddings and compute cosine similarities between the image-audio embeddings (denoted $s(i,a)$) and text-audio embeddings (denoted $s(t,a)$). We use the weighted sum $\lambda_1 s(i,a)+\lambda_2 s(t,a)$ to retain the audio files corresponding to the top-N similarity scores. We repeat this process for all images in each mood category to curate our audio-design dataset.

Our proposed MELON dataset consists of 488,510 images and corresponding captions and 7,737 music audios belonging to 50 moods/themes. Each mood category consists of $\approx 10k$ images. A distribution plot of the audio samples per category as well as per-mood word clouds to demonstrate the diversity and variability of our dataset are provided in the supplementary material\footnote{Link to supplementary document: \url{https://shorturl.at/hsDU5}. We plan to make materials publicly available at this link.}. In Table \ref{tab:datasets}, we quantitatively compare the proposed MELON dataset with existing datasets. One can note that while existing datasets are focused on natural images and videos, our proposed dataset is unique in the sense it contains creative illustrations, vectors, templates, and background designs with complete descriptions of the image content as part of the caption. With $\approx 500k$ images, our dataset will help the community build robust models for both music retrieval (MR) and classification (C) tasks.

\section{Multi-modal Cross-attention  Audio Retrieval (MMCAR) }
\label{sec:mmcar}
Here, we describe our proposed algorithm for retrieving audio files given an input design document. Our key algorithmic novelty is a multi-modal cross-attention module that operates on feature vectors from all three input modalities (image, text, and audio) and learns a common representation space for the downstream retrieval task.

Figure~\ref{fig:blockdiagram} visually summarizes our proposed algorithm. During training, given input triplets from our MELON dataset, we first use per-modality embedding extractors to compute feature vectors $\mathbf{i}$, $\mathbf{t}$, and $\mathbf{a}$ for the image, text, and audio modalities respectively. For $\mathbf{i}$ and $\mathbf{t}$, we use the CLIP \cite{radford2021clip} model to obtain 512-dimensional embeddings each. For $\mathbf{a}$, we train a Resnet-18 model (for audio classification) on the publicly available VGGSound dataset.

Given the $\mathbf{i}$, $\mathbf{t}$, and $\mathbf{a}$ embeddings, we propose a multi-modal cross-attention operation to learn a unified multi-modal design-audio embedding. Given the three feature vectors, we perform pairwise cross-attention pooling taking any two modalities $\x,\y\in \{\boldsymbol{i},\boldsymbol{t},\boldsymbol{a}\}$ such that $\x \in \mathcal{R}^{d}$ is the query  and $\y \in \mathcal{R}^{d}$ is the key, resulting in an output $\hat{x}$. Similarly, by interchanging $\x$ and $\y$, we compute the output $\hat{y}$. These two outputs are then used to compute a common embedding $\mathbf{u}_{xy}$ for this particular pair of $x$ and $y$. We repeat this for all the possible pairs ($(x=i, y=t), (x=i, y=a), (x=t, y=a)$), and use the corresponding outputs to obtain the proposed unified embedding $\mathbf{u}_{\text{all}}$ as:
 \begin{equation}
\label{eqn:crossattention}
\begin{split}
\C_{xy}&= \x\y^T  \in \mathcal{R}^{dXd} \quad and \quad \C_{yx}= \y\x^T  \\ 
\Sb_x &= \sigma(\C_{xy} *\W+\B) \in \mathcal{R}^{dXd} \\
\hat{\x}&= diag(\Sb_x\C_{xy}^T)\\
\end{split}
\end{equation}
%  \text{Similarly, }
 \begin{equation}
 \begin{split}
\Sb_y &= \sigma(\Sb_{yx} *\W+\B) \in \mathcal{R}^{dXd} \\
\hat{\y}&= diag(\Sb_y\C_{yx}^T) \\
\end{split}
 \end{equation}
  \begin{equation}
 \begin{split}
  \boldsymbol{u}_{xy}&=\hat{\x} \oplus \hat{\y}  \in \mathcal{R}^{2dX1}\\
\boldsymbol{u}_{all}&=\boldsymbol{u}_{it} \oplus \boldsymbol{u}_{ia}  \oplus  \boldsymbol{u}_{ta} \in \mathcal{R}^{6dX1}
\end{split}
 \end{equation}
 
This unified embedding $\mathbf{u}_{\text{all}}$ is then passed to a fully connected neural network unit which generates, with a sigmoid operation, a scalar score $\hat{z}$ in range $\in [0,1]$. We compare this with the ground truth score $z=1$ (if the input is a correct pair) and $z=0$ otherwise, resulting in a binary cross-entropy training objective
\vspace{-10pt}
\begin{equation}
 L=\frac{1}{B}\sum_{i=1}^{B}z\text{log}(\hat{z})+(1-z)\text{log}(1-\hat{z})
\end{equation}
where B is the batch size. During inference, given a design document input and a database of $n$ audio samples from which to retrieve relevant files, our model computes the similarity scores $\hat{z}_1, \hat{z}_2,...,\hat{z}_n$ for the input image-text pair with all the $n$ audio samples. Given these scores, we pick the audio files corresponding to the top-$k$ highest scores as the retrieval results (see Fig~\ref{fig:blockdiagram} (right)).

\begin{figure*}
\centering
\includegraphics[width=0.8\textwidth, trim={0cm 2cm 2.5cm 4.5cm},clip]{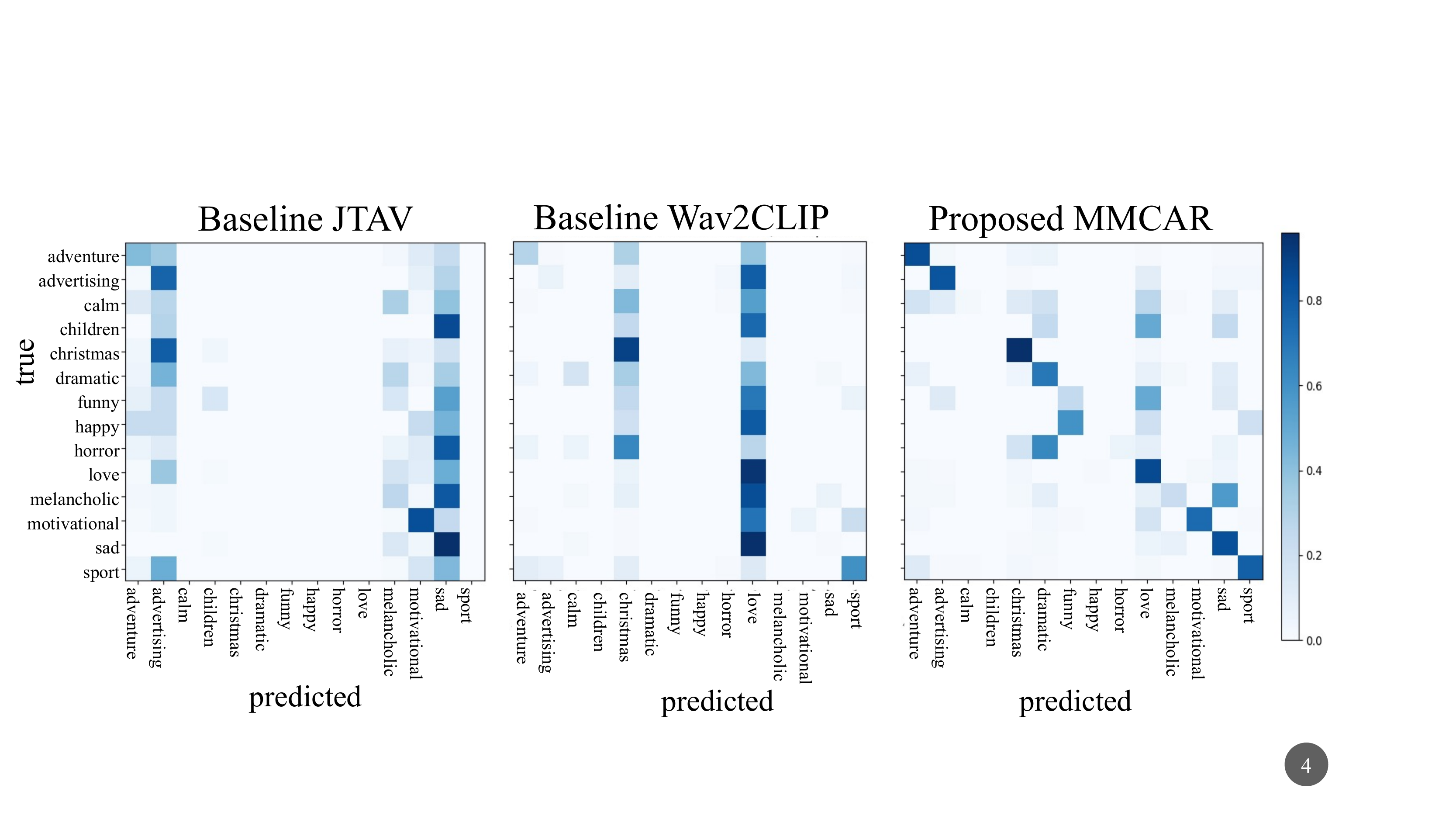}
\caption{Confusion matrix of proposed MMCAR vs. baseline JTAV and Wav2CLIP.}
\label{fig:confusionmatrix}
\end{figure*}

\begin{table}[]
\scalebox{0.9}{
\begin{tabular}{|l|l|l|l|l|}
\hline
adventure & advertising & drama       & funny        & love       \\
fun       & commercial  & dramatic    & groovy       & romantic   \\
game      & corporate   & movie       & happy        & nature     \\
holiday   & ambiental   & dream       & hopeful      & summer     \\
horror    & calm        & emotional   & motivational & retro      \\
space     & relaxing    & heavy       & melodic      & background \\
sport     & soft        & melancholic & children     &            \\
upbeat    & mellow      & sad         & christmas    &            \\ \hline
\end{tabular}
}
\caption{Lists of moods/themes used for training and evaluation.}
\label{tab:moods}
\vspace{-6pt}
\end{table}

\begin{figure}
\centering
\includegraphics[width=0.4\textwidth, trim={0cm 0cm 11cm 0cm},clip]{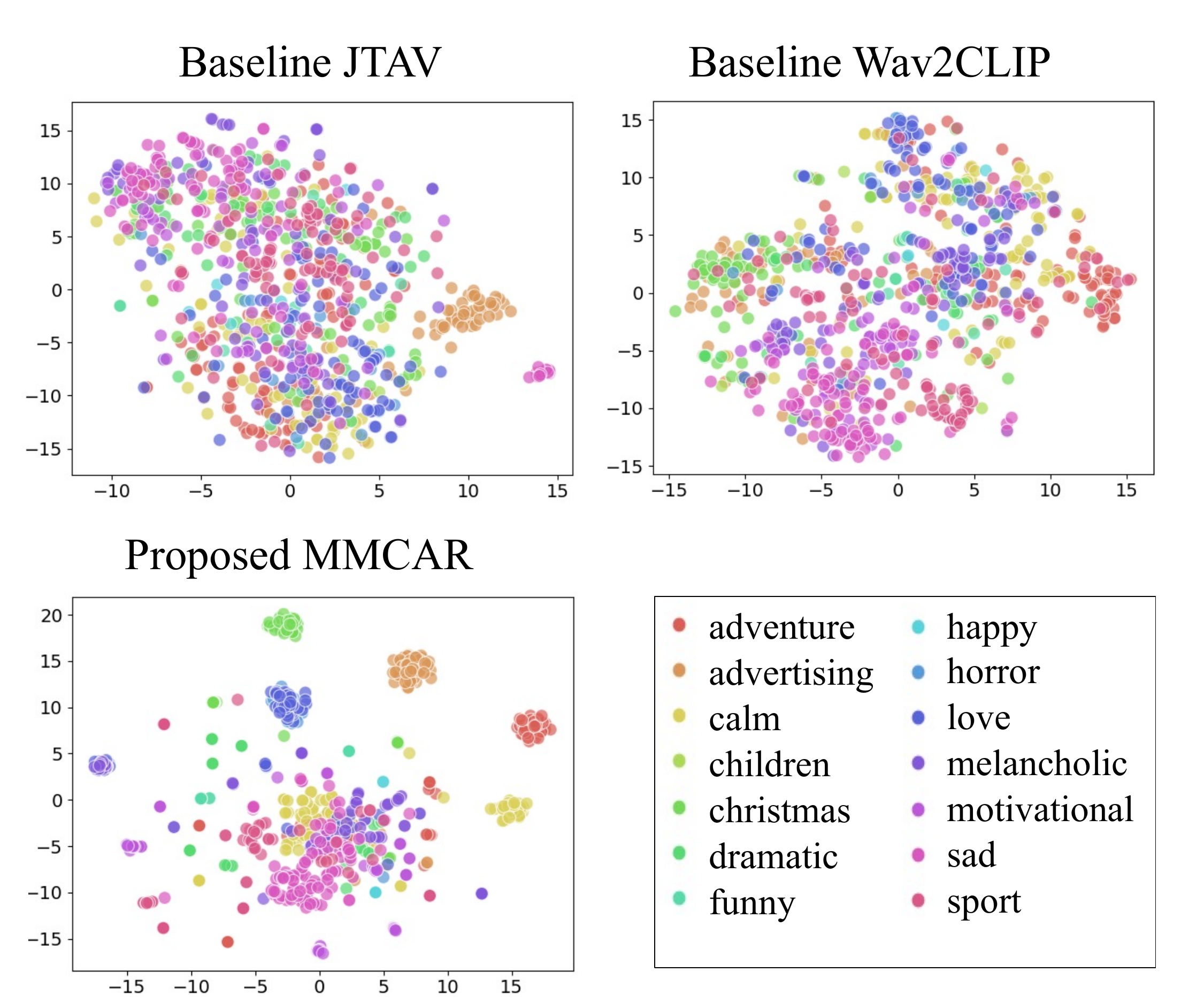}
\caption{t-SNE plots of baselines vs. proposed approach.}
\label{fig:tsne}
\vspace{-8pt}
\end{figure}
\vspace{-5pt}

\section{Experiments and Results}
Since our proposed problem, data, and algorithm is centered around audio retrieval given image-text design documents, the closest baselines in the literature include JTAV \cite{liang2018jtav} and Wav2CLIP \cite{wu2022wav2clip}. While JTAV operates on an image, its caption, and the textual lyrics of an audio to learn features, Wav2CLIP uses the CLIP \cite{radford2021clip} image encoder and an audio autoencoder to map audio and image features close.

To benchmark the performance of these algorithms and compare them to our proposed method on our new dataset, we use rank-based evaluation metrics proposed in prior work \cite{liang2018jtav}. In particular, we use the \textit{Med r} $\in[1, M]$ metric that represents the medium rank of the ground-truth retrieved audio, where $M$ is the maximum number of classes considered. A lower value of \textit{Med r} indicates better performance. We also use recall@k ($k=\{1,5,10,15,20\}$) that is the fraction of ground-truth audios retrieved in the top-k ranked items across all test cases, and higher values indicate better performance.

For training and evaluating all models, we selected 38 moods in our dataset based on maximum uniqueness among all the 50 moods (see Table~\ref{tab:moods} for a list). We construct the training split using $60\%$ of the image-caption pairs and audio samples. Each triplet comprises the image, caption, and audio along with a $1/0$ label based on the correct audio mapping. To form positive triplets, we select $10$ audio samples for each design based on mood as discussed in Section \ref{sec:pairformation}. The negative triplets are formed by randomly selecting $5$ different moods from the mood of the image and selecting $2$ audio samples from each mood. The remaining $40\%$ of the data is equally split to obtain validation and test splits with $20\%$ of the samples each.

We next present our evaluation results. Since we seek to retrieve the best matched audio file in terms of mood, we first compute a mean feature vector for each mood from our audio repository. Then, given the feature vector for an input design document, we generate a score vector $\boldsymbol{\hat{z}}\in\mathcal{R}^{M}$, where $M=38$ as noted above, for the mean features of all the audio files. We then compute the recall@k and \textit{Med r} metrics using the reference mood label.% for each design document in the validation and test sets. 

In Table~\ref{tab:results}, we compare the performance of our proposed MMCAR algorithm with JTAV and Wav2CLIP baselines on both val and test sets. One can note that our proposed MMCAR gives the lowest $Med$ $r$ of $3.9$ and the highest recall accuracy at all ranks, e.g., $79\%$ accuracy at k=5 on the test set compared to JTAV's $30\%$ and Wav2CLIP's $39\%$, accounting for more than $100\%$ relative improvements. For an even more fair comparison, we also reimplemented our MMCAR with Wav2CLIP's features (noted MMCAR* in table). While MMCAR* leads to performance degradation when compared to our original, end-to-end-learned MMCAR model, it is still substantially better than the baseline Wav2CLIP approach. This provides evidence for our multimodal cross-attention module's discriminative capabilities in the shared image-text-audio feature space.
%Using Wav2CLIP audio features instead of Resnet features leads to degradation in performance. However, it is better than the Wav2CLIP baseline approach, which does not incorporate cross-attention. This shows that the cross-attention module helps in the task. Cross-attention helps capture the essential features from the 300D features of all the modalities by using cross modalities which helps in extracting the needed information to identify the mood of the audio. 

To provide additional evidence, we show t-SNE plots of the learned design document embeddings in Figure~\ref{fig:tsne} where one can see a clearer clustering, compared to baseline approaches, of the features according to the mood using the proposed MMCAR approach. In Figure~\ref{fig:confusionmatrix}, we compare MMCAR's confusion matrix with the baseline ones for a random selection of 14 moods, where one can note while baseline predictions are biased towards a few specific moods, the proposed method has a close-to-diagonal matrix as expected.

\begin{table}[]
\scalebox{0.9}{
\begin{tabular}{l|l|lllll}
\hline
\multirow{2}{*}{Methods} & \multirow{2}{*}{$Med$ $r$} & \multicolumn{5}{c}{Recall@k}                                                                                       \\ \cline{3-7} 
                         &                        & \multicolumn{1}{l|}{k=1} & \multicolumn{1}{l|}{k=5} & \multicolumn{1}{l|}{k=10} & \multicolumn{1}{l|}{k=15} & k=20 \\ \hline
  \multicolumn{7}{c}{Val}  \\   \hline
JTAV \cite{liang2018jtav}      & 17.1                       & \multicolumn{1}{l|}{0.12}    & \multicolumn{1}{l|}{0.30}    & \multicolumn{1}{l|}{0.35}     & \multicolumn{1}{l|}{0.46}     & 0.62      \\
Wav2CLIP \cite{wu2022wav2clip}  & 9.2                       & \multicolumn{1}{l|}{0.12}    & \multicolumn{1}{l|}{0.37}    & \multicolumn{1}{l|}{0.66}     & \multicolumn{1}{l|}{0.83}     &0.92      \\
\textbf{MMCAR [Ours] }           & \textbf{3.8} & \multicolumn{1}{l|}{\textbf{0.42}} & \multicolumn{1}{l|}{\textbf{0.80}} & \multicolumn{1}{l|}{\textbf{0.92}} & \multicolumn{1}{l|}{\textbf{0.96}} & \textbf{0.98} \\
MMCAR* & 7.0 & \multicolumn{1}{l|}{0.23} & \multicolumn{1}{l|}{0.59} & \multicolumn{1}{l|}{0.77} & \multicolumn{1}{l|}{0.87} & 0.93 \\ \hline
\hline
 \multicolumn{7}{c}{Test}  \\   \hline
JTAV \cite{liang2018jtav}      &16.9                        & \multicolumn{1}{l|}{0.12}    & \multicolumn{1}{l|}{0.30}    & \multicolumn{1}{l|}{0.35}     & \multicolumn{1}{l|}{0.45}     &0.62      \\
Wav2CLIP \cite{wu2022wav2clip}  & 9.2                        & \multicolumn{1}{l|}{0.11}    & \multicolumn{1}{l|}{0.39}    & \multicolumn{1}{l|}{0.67}     & \multicolumn{1}{l|}{0.83}     &0.91      \\
\textbf{MMCAR [Ours]}            &\textbf{3.9} &\multicolumn{1}{l|}{\textbf{0.42} } & \multicolumn{1}{l|}{\textbf{0.79}} & \multicolumn{1}{l|}{\textbf{0.92}} & \multicolumn{1}{l|}{\textbf{0.96}} &\textbf{0.98}  \\
MMCAR* &7.2 & \multicolumn{1}{l|}{0.22} & \multicolumn{1}{l|}{0.56} & \multicolumn{1}{l|}{0.76} & \multicolumn{1}{l|}{0.89} &0.93  \\ \hline
\end{tabular}
}
\caption{Proposed MMCAR vs. baselines.}
\label{tab:results}
\vspace{-12pt}
\end{table}
\vspace{-6pt}
\section{Summary}
We considered and proposed a new problem of retrieving relevant audio files given multimodal design documents as input. In the absence of any relevant datasets in the literature, we built and presented a first-of-its-kind multimodal design-audio dataset called MELON comprising hundrends of thousands of design files with mapped audio files. We then proposed a multimodal cross attention algorithm that enables training neural networks to learn a joint image-text-audio feature space for design documents and used it to retrieve relevant audios given a certain design input at test time. We benchmarked our algorithm against the existing state of the art on our new dataset and hope that this will spur further research in this area.

%In this paper, we aim to address the task of music retrieval for multimodal design documents using multimodal image-caption pair. We have proposed the ``MELON" dataset, a one-of-a-kind multimodal dataset, to address this problem. It can be used for various retrieval tasks, including music retrieval and other classification tasks, e.g. image mood classification. The proposed dataset has been benchmarked with state-of-the-art approaches. We have developed model architectures to accept image, text and audio inputs and learn the essential features using cross-attention to retrieve the best music for the given image by identifying the correct mood/theme of the input. The experimental results have shown the effectiveness of the proposed model over the baseline models.
\vspace{-6pt}
\section{ACKNOWLEDGEMENTS}
\vspace{-2pt}
 The authors would like to thank Dr.~Sriram Ganapathy of LEAP Lab, Indian Institute of Science, Bangalore, for his valuable input and help in offering the required resources to run the experiments.

% Below is an example of how to insert images. Delete the ``\vspace'' line,
% uncomment the preceding line ``\centerline...'' and replace ``imageX.ps''
% with a suitable PostScript file name.
% -------------------------------------------------------------------------

%\vfill
%\pagebreak

\bibliographystyle{IEEEbib}
\bibliography{Template}

\end{document}